\documentclass[aip,jap,amsmath,amssymb,preprint]{revtex4-1}

\usepackage{graphicx}
\usepackage{dcolumn}
\usepackage{bm}

\begin{document}

\preprint{AIP/123-QED}

\title[]{The electronic properties of a core/shell/well/shell spherical quantum dot with and without a hydrogenic impurity}

\author{Hatice Ta\c{s}}
\email{h.tas@live.com}
\affiliation{Department of Physics, Faculty of Science, Sel\c{c}uk University, Campus 42075 Konya, Turkey}

\author{Mehmet \c{S}ahin}
\email{mehmet.sahin@agu.edu.tr}
\affiliation{Department of Material Science and Nanotechnology Engineering, Abdullah G\"{u}l University, Kayseri, Turkey}
\affiliation{Department of Physics, Faculty of Science, Sel\c{c}uk University, Campus 42075 Konya, Turkey}

\date{\today}

\begin{abstract}
In this study, we have performed a detailed investigation of the electronic properties of a core/shell/well/shell multi-layered spherical quantum dot, such as energy eigenvalues, wave functions, electron probability distribution and binding energies. The energy eigenvalues and their wave functions of the considered structure have been calculated for cases with and without an on-center impurity. For this purpose, the Schr\"{o}dinger equation has been numerically solved by using the shooting method in the effective mass approximation for a finite confining potential. The electronic properties have been examined for different core radii, barrier thickness and well widths in a certain potential. The results have been analyzed in detail as a function of the layer thickness and their physical reasons have been interpreted. It has found that the electronic properties are strongly depending on the layer thickness.
%
\end{abstract}

\pacs{73.21.La, 73.22.-f, 73.63.Kv}
\keywords{Multi-shell quantum dot, hydrogenic impurity, binding energy}
\maketitle

\section{Introduction}

The recent developments in the manufacturing and growth technology have caused the fabrication of the low dimensional systems which are called nanostructures. These structures are different from the macroscopic structures and they present rather interesting properties.\cite{fer} Because of these properties, the nanostructures have a wide area of the research and interest for the scientists.\cite{lin} The electron motion can be confined in all three dimensions with advanced manufacturing technology and so, these improvements have made possible the production of quantum dots (QDs). These confinements have led to sharp energy levels and density of states as which were found in atoms. Because of that, QDs are also called artificial or super atom.\cite{xia0,ser,wag,tul,bim} Outstanding properties of QDs are changes in energy spectrum by doing some modifications in systems which have different geometry and formed via growth technology. The investigations claim that the electronic properties of QDs are seriously connected to their size, shape and dimensions.\cite{sar} Because of the possibility to control of the  sizes, shapes, energy levels and electron number of QDs, there has been considerable amount of theoretical and experimental studies on these structures.

As is well known, an impurity atom in a semiconductor is a different one from the semiconductor's atom and it has an additional electron more than that required to make a chemical bonds with neighboring atoms. This more electron can be easily donating to the host material. Such impurity atoms are known as donors. If a donor impurity has an additional one electron and is unionized, this kind of impurity is known as hydrogenic one because of that it is very similar to a hydrogen atom. Thus, most of properties of this impurity are overlap with that of a hydrogen atom. An example of a typical hydrogenic donor impurity in a GaAs crystal is silicon.\cite{har} More detail related with a hydrogenic donor impurity can be found in Ref. \onlinecite{har}.

The existence of impurities causes some changes in the performance of quantum devices and their electronic, optical and transport properties because of the Coulomb interaction between the electron and the impurity atom. Although it does not provide fully accurate descriptions about the QDs, the study on hydrogenic impurity problem in semiconductor QDs is very useful model, for understanding the optical and the electronic properties of these structures. Therefore, this model is used in a number of studies, since many physical properties of an impurity can be explained in these kinds of heterostructures, even though it is simple and a somewhat rough estimate.\cite{sah1,sah2,zhu,por,joh} The confinement in all three dimensions in QDs causes a reduction of the distance between the electron and the impurity leading an increase in the Coulomb interaction. The hydrogenic impurity states in the QDs have been widely studied by many authors due to their importance for potential applications in optoelectronic and photonic devices.  Zhu at al.\cite{zhu} and Porras-Montenegro and Perez-Merchancano\cite{por} have carried out pioneering studies about confinement effect on hydrogenic impurity states in QDs. In these studies, binding energy has been probed as a function of QD's size and impurity location. In addition to these studies, the hydrogenic donor binding energy has been surveyed as theoretically by many authors for QDs which have different geometries such as spherical, cylindrical, pyramidal and rectangular.\cite{joh,elm,bet,yua,oli,alv,bos0} Riberio and Latge\cite{rib} worked on a cubic quantum heterostructures and they have showed that the binding energy values are close to those of cubic and spherical QDs which have near volume. Bose and Sarkar\cite{bos1} have used perturbation and variation technique to obtain hydrogenic donor binding energies in a spherical QD for an infinite parabolic confinement. The study of the impurity effect on the binding energy has made possible the fabrication of a low threshold laser diode.\cite{gro} Jayam\cite{jay} has investigated the effect of the electric field on a hydrogenic donor binding energy and same problem has been studied by Peter \cite{pet} for the the effect of a hydrostatic pressure. Hydrogenic donor binding energy for excited states in a spherical QD has been analyzed by Sadeghi.\cite{sad}

Recent improvements in growth technology have resulted in manufacturing of multi-layered spherical quantum dots (MSQD) and so, MSQD has been studied theoretically by many researchers.\cite{hsi} The binding energy of a hydrogenic impurity in a GaAs-AlGaAs MSQD has been calculated as a function of barrier and the inner dot thickness for different potentials.\cite{akt} For the same structure, Boz et al.\cite{boz} explored effects of the geometry on hydrogenic impurity energy states by using a fourth order Runge-Kutta method.

In this paper, we investigate the electronic properties of a core/shell/well/shell QD structure for both ground (1s) and excited (1p) states. We compare these properties for cases with and without the impurity and for different layer thickness. At the same time, the effect of the variations in the geometry of the structure on the donor binding energy is studied. All results are presented for both cases with and without the impurity.

The rest of the paper is organized as follows: In the next section, we describe the model and calculations briefly. Results and discussion are presented as detailed in section III. In the last section, a short conclusion is given.

\section{Model and Calculations}

\begin{figure}
\includegraphics[width=3.4in]{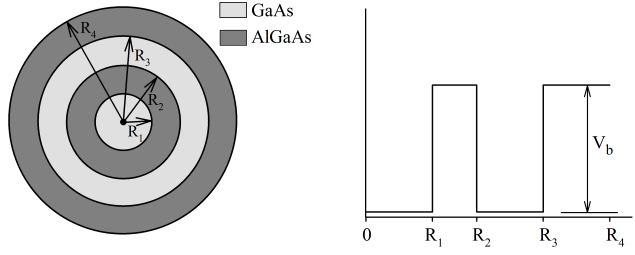}
\caption{\label{fig:1} Schematic representation of a core/shell/well/shell QD and its potential profile.}
\end{figure}

In this study, we have considered a spherical core/shell/well/shell QD containing a single electron. This structure comprises of two GaAs dots one embedded in the other. Also, in this structure, each of QDs is isolated by means of AlGaAs which provides a finite confining potential for the electron. Out of this structure has been coated by AlGaAs again as seen in Fig.\ref{fig:1}. In our system, spherical QDs have been interacted through potential barrier which allows electrons to tunnel between the core and well regions. The shell thickness is $T_s=R_2-R_1$ and the well width is $T_w=R_3-R_2$. The effective mass approximation and BenDaniel-Duke boundary conditions are used in the calculations. Here, the impurity position is taken as center of the structure. In the effective mass approximation, the single-particle Schr\"{o}dinger equation is given as

\begin{widetext}
\begin{equation}
\label{eq1}
\left[ { - \frac{{\hbar ^2 }}{2}\vec \nabla _r \left(
{\frac{1}{{m^* \left( r \right)}}\vec \nabla _r } \right) -
\frac{{Ze^2 }}{r} + \frac{{\ell \left( {\ell  + 1} \right)\hbar^2 }}{{2m^* \left( r \right)r^2 }}+ V_b} \right]R_{n,\ell } \left( r \right) = E _{n,\ell} R_{n,\ell } \left( r \right).
\end{equation}
\end{widetext}

\noindent
Here, $\hbar$ is reduced Planck constant, $m^{\ast}(r)$ is the position-dependent electron's effective mass, $Z$ is the charge of the impurity, $\ell$ is the angular momentum quantum number, $V_{b}$ is the finite confining potential, $E_{n,\ell}$ is the electron energy eigenvalue for a specified principle quantum number ($n$) and angular momentum quantum number ($\ell$), $R_{n,\ell}(r)$ is the radial wave function of the electron. It should be noted that $Z=0$ and $Z=1$ correspond to cases without and with a hydrogenic donor impurity, respectively. The mathematical expression of the confining potential is

\begin{equation}
 \label{eq2}
 V_b=\left\{{\begin{array}{l}
 0,\,\,\,\,\,\,\,\,\,\,\,\,\,\,\,\,\,\,\,\,\,\,\,\,r \leq R_1 \,\,\, $and$ \,\,\, R_2\leq r \leq R_3\\\\
 228 \,\, $meV$,\,\,\,\,R_1<r<R_2 \,\,\, $and$ \,\,\, r>R_3 \\
 \end{array}} \right. .
\end{equation}

\noindent
Here, 228 meV is the conduction band offset in GaAs/AlGaAs junction.\cite{por,ada}

In order to determine the single particle energy levels and corresponding wave functions, Eq.(\ref{eq1}) is solved full numerically. Shooting method is used for the determination of the single particle energies and corresponding wave functions. As is well known, this technique converts an eigenvalue problem to an initial value problem. For this purpose, Hamiltonian operator is discretized on a uniform radial mesh in 1D using the finite differences, then Eq.(\ref{eq1}) can be reduced to an initial value equation by means of

\begin{widetext}
\begin{equation}
\label{eq3}
R_{n,\ell}(i+1)=\left(\frac{r}{r+h}\right)\left[2+\frac{2m^*h^2}{\hbar^2}\left(V(i)+
\frac{\ell(\ell+1)\hbar^2}{2m^*r^2}-E _{n,\ell}-\frac{Ze^2}{\varepsilon r}\right)\right]R_{n,\ell}(i)-\left(\frac{r-h}{r+h} \right)R_{n,\ell}(i-1),
\end{equation}
\end{widetext}

\noindent
where $i$ is the index of mesh points, $h$ is the distance between two mesh points and it is chosen as 0.005. The details of this method can be found in Ref. \onlinecite{har}.

The neutral donor binding energy depends on position of the impurity and geometry of the structure \cite{xia1} and is determined as

\begin{equation}
\label{eq4}
E_b=E_{n,\ell}-E_{n,\ell}(D^0),
\end{equation}

\noindent
where $E_{n,\ell}$ and $E_{n,\ell}(D^0)$ are the energy levels for cases without and with the impurity, respectively.

\section{Results and Discussion}

The atomic units have been used throughout the calculations, where Planck constant $\hbar=1$, the electronic charge $e=1$ and the bare electron mass $m_0=1$. Effective Bohr radius is $a_0=0.529\frac{\kappa_1}{m_1^*}\simeq100${\AA} and effective Rydberg energy is $R_y=13.6\frac{m_1^*}{\kappa_1^2}\simeq5.25$ meV\cite{por,buc}. The material parameters have been taken as\cite{ada,ada1} $m_{GaAs}$=0.067$m_0$, $m_{AlGaAs}$=0.088$m_{0}$, $\kappa_{GaAs}$=13.18 and $\kappa_{AlGaAs}$=12.8. Also the effective masses of electrons inside GaAs and AlGaAs are $m_1^*$ and $m_2^*$, and the dielectric constants are $\kappa_1$ and $\kappa_2$, respectively. The position-dependent effective mass and the dielectric constant may be defined as follows\cite{buc}

\begin{eqnarray}
\label{eq5}
\nonumber
m^\ast(r) = \left\{ {\begin{array}{l}
 1,\,\,\,\,\,\,\,r \leq R_1 \,\,\, $and$ \,\,\, R_2\leq r \leq R_3 \\\\
 \frac{m_2^\ast }{m_1^\ast },\,\, R_1<r<R_2 \,\,\, $and$ \,\,\, r>R_3 \\
 \end{array}} \right. \\ \nonumber
\\
\kappa(r) = \left\{ {\begin{array}{l}
 1,\,\,\,\,\,r \leq R_1 \,\,\, $and$ \,\,\, R_2\leq r \leq R_3 \\\\
 \frac{\kappa _2 }{\kappa _1 },\,\,R_1<r<R_2 \,\,\, $and$ \,\,\, r>R_3 \\
 \end{array}} \right. .
\end{eqnarray}

\subsection{Effect of the layer thickness on the electronic properties}

\begin{figure}
\includegraphics[width=2.5in]{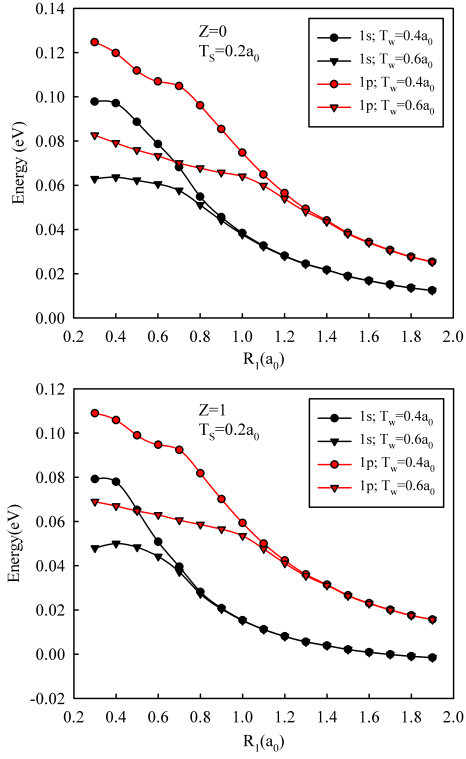}
\caption{\label{fig:2} Ground and excited state energy values as a function of core radius ($R_1$) for constant shell thickness ($T_s=0.2a_0$) and two different well widths ($T_w=0.4$ and $0.6a_0$) for $Z=0$ (top panel) and $Z=1$ (bottom panel).}
\end{figure}

\begin{figure}
\includegraphics[width=3.0in]{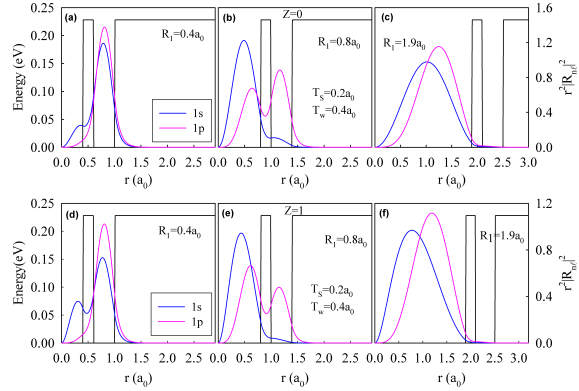}
\caption{\label{fig:3} The electron density of ground and excited states for $Z=0$ (top panel) and $Z=1$ (bottom panel). The thickness of the layers are specified on the figures.}
\end{figure}

Figure \ref{fig:2} shows the variation of the energy levels for $Z=0$ (top panel) and $Z=1$ (bottom panel) as a depending upon $R_1$ for both ground ($1s$) and excited ($1p$) states in a MSQD. In this calculation, the barrier thickness is taken as $T_s=R_2-R_1=0.2a_0$ and the results are comparatively given for two different well widths specified on the figures. As seen from the figures, when $R_1$ increases, both ground and excited state energies exhibit decreasing tendency. This is an expected result because of the relation between energy and radius, $E\propto1/R_1$. Also, we observe that ground and excited state energies have different values until certain $R_1$ value for different well widths in both $Z=0$ and $Z=1$ cases. After this $R_1$ value, ground states energies close to each other and they have same values with further increasing of the core radius. At the same time, the figures show that these consequences are valid for also excited states. In $Z=0$ case, while ground state energies have same values at around $R_1=1.0 a_0$, these changes are at about $R_1=1.4 a_0$ for excited state ones. In case of $Z=1$, this behaviors are observed at approximately $R_1=0.7 a_0$ and $R_1=1.2 a_0$ for ground and excited states, respectively. This is because, in small value of the core radius, probability of ground and excited state electrons  in the well region is larger than that in the core region as seen from Fig. \ref{fig:3}. When the core radius increases, the electron probability starts to increase in this region due to the rule of energy minimum tendency and relation between energy and core radius as mentioned above. This increase in the probability is more evident for case with $Z=1$ due to attractive Coulomb potential of the impurity as seen in Fig. \ref{fig:3} (e). After the specified core radii, the probability densities of both ground and excited states with and without the impurity become maximum in the core region as seen in Fig. \ref{fig:3} (c) and (f). Because, the electron and energy levels do not feel the effect of the well region. As a result, core/shell/well/shell structure exhibits the properties of a single QD with a finite potential barrier.

\begin{figure}
\includegraphics[width=2.5in]{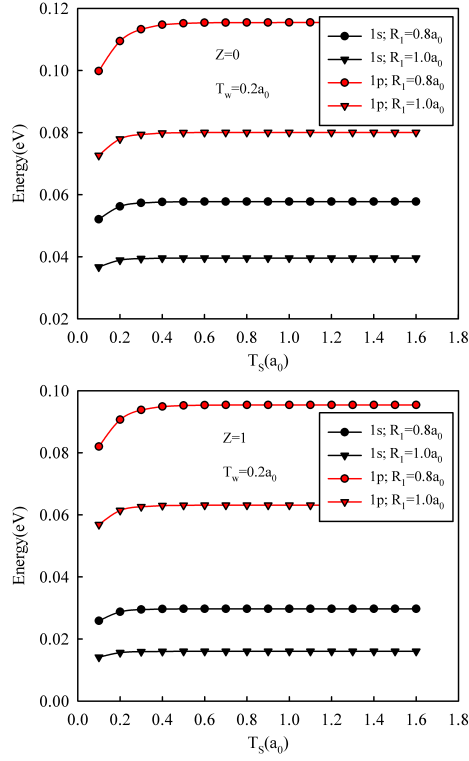}
\caption{\label{fig:4} Variation of ground and excited state energy values with shell thickness ($T_s$) for constant well width ($T_w=0.2a_0$) and two different core radii ($R_1=0.8$ and $1.0a_0$) for $Z=0$ (top panel) and $Z=1$ (bottom panel).}
\end{figure}

\begin{figure}
\includegraphics[width=3.0in]{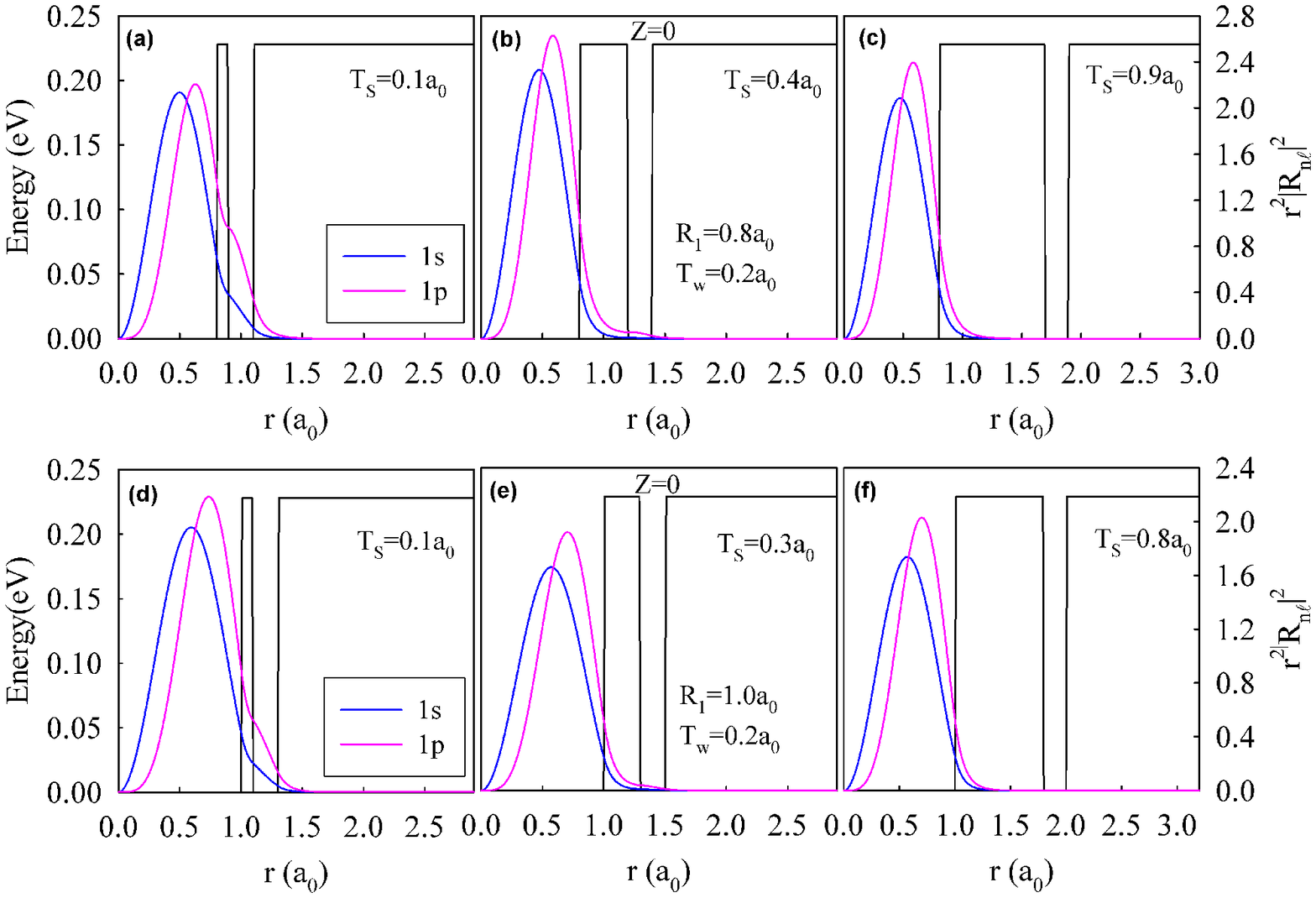}
\caption{\label{fig:5} The electron density of ground and excited states for $Z=0$ and two different core radii, $R_1=0.8a_0$ (top panel) and $R_1=1.0a_0$ (bottom panel). The thickness of the layers are specified on the figures.}
\end{figure}

In the system under consideration, AlGaAs layer embedded between two GaAs QDs creates a finite potential barrier and so, the electron tunneling occurs between the core and well regions. Thus, it can be expected that the variation of the barrier thickness affects the energy values of this system for cases with and without the impurity. In order to understand these variations, we must analyze Fig. \ref{fig:4} which shows the variation of the energies of both ground and excited state energies with the shell thickness for $Z=0$ (top panel) and $Z=1$ (bottom panel). The figures are plotted for two different core radii ($R_1=0.8$ and $1.0 a_0$) and same well widths ($T_w=0.2 a_0$). As seen from these figures, in all cases, while the energy values increase until a small fraction of the shell thickness, they remain constant values after those thickness. These tendencies can be explained as follows: When the barrier thickness is small, the wave functions of ground and excited states can tunnel between the core and well region for both $Z=0$ and $Z=1$ cases as seen in Fig. \ref{fig:5}. As a result of this, the electron which is in the core region (or in the well region) is affected by the well region (or core region) for both ground and excited states. This response is less evident for $Z=1$ case because of attractive Coulomb potential of the impurity. However, an increase in of the barrier thickness decreases possibility of the tunneling between core and well regions and so, the electron confined in the core region can not sense the effect of the well region. Because of that, the energies have same value in the case of further increasing of barrier thickness. We conclude that the barrier thickness is rather important until a critical value for tuning the electronic structure of a MSQD.

\begin{figure}
\includegraphics[width=2.5in]{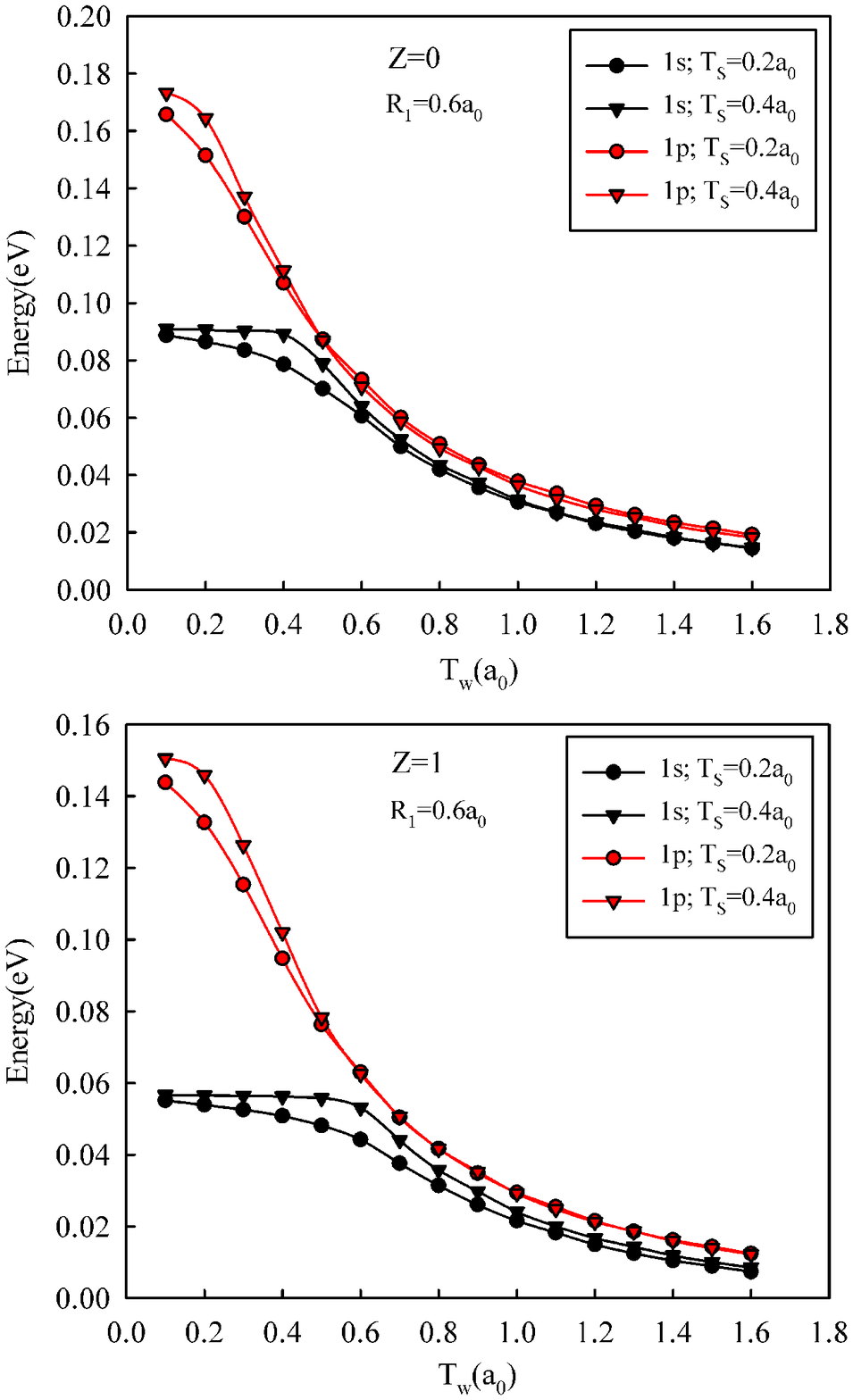}
\caption{\label{fig:6} Variation of ground and excited state energy values as a function of well width ($T_w$) for constant core radius ($R_1=0.6a_0$) and two different shell thickness ($T_s=0.2$ and $0.4a_0$) for $Z=0$ (top panel) and $Z=1$ (bottom panel).}
\end{figure}

\begin{figure}
\includegraphics[width=3.0in]{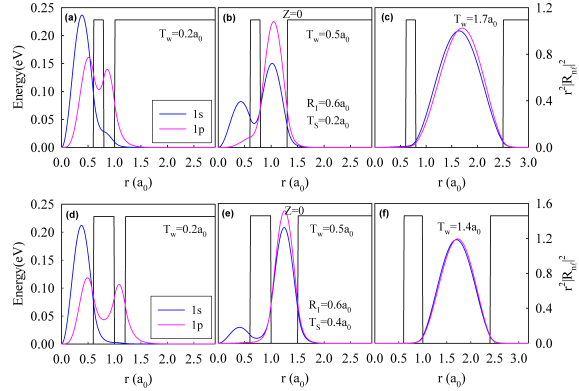}
\caption{\label{fig:7} The electron density of ground and excited states for $Z=0$ and two different well width, $T_s=0.2a_0$ (top panel) and $T_s=0.4a_0$ (bottom panel). The thickness of the layers are specified on the figures.}
\end{figure}

Fig. \ref{fig:6} shows the variation of the ground and excited state energy values with the well width ($T_w$) for $Z=0$ (top panel) and $Z=1$ (bottom panel). The figures are plotted for two different barrier thicknesses ($T_s=0.2$ and $0.4a_0$) and the core radius is taken as $R_1=0.6a_0$. As seen from the figure, the energy levels, especially 1s states, remain constant until $T_w=0.4a_0$ for case without impurity, when $T_s=0.4a_0$. This treatment continues to approximately $T_w=0.6a_0$ for case with impurity because of the attractive Coulomb potential of the impurity. These changes are more smooth for $T_s=0.2a_0$, because the electron is localize in the core region (or in the well region), it can leak easily to the well region (or to the core region). These treatments can be obviously observed in Fig. \ref{fig:7}. In the both cases of $Z=0$ and $Z=1$, the difference between ground and excited state energies is considerably large for the small well widths. All the energies decrease with further increase in the well widths. As is well known, when the well width increases, the quantum confinement effect decreases. Therefore, the consecutive energy levels become close to each other in both $Z=0$ and $Z=1$ cases. In addition, as seen from Fig. \ref{fig:7}, the electron escapes from the core region and it is completely confined in the well region with increasing well widths. If the well region is much more extended, the electron feels less confinement effect on it and hence, the energy values of the electron go towards a free particle ones. Moreover, ground and excited states energies for $Z=1$ case are smaller than those for $Z=0$ case. Also, the difference between ground and excited states energies in $Z=1$ case is larger than those for $Z=0$ case. All of these situations are originated from the attractive Coulomb interaction between the electron and impurity.

\subsection{Effect of the layer thickness on the donor binding energy}

\begin{figure}
\includegraphics[width=2.5in]{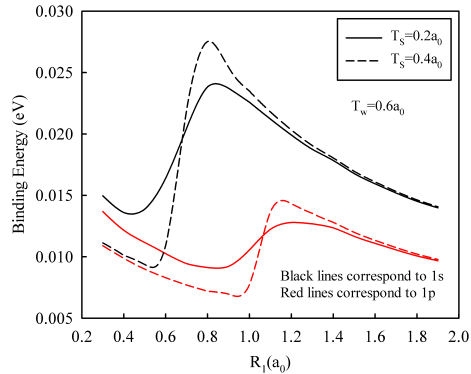}
\caption{\label{fig:8} Variation of ground and first excited state impurity binding energies with core radius ($R_1$) for $T_w=0.6 a_0$ and two different shell thickness, $T_s=0.2$ and $0.4 a_0$.}
\end{figure}

Variation of ground and first excited state binding energies of the donor impurity as a function of the core radius ($R_1$) is given in Fig. \ref{fig:8} for constant well width and two different shell thickness that are specified on the figure. As seen from the figure, ground state binding energy exhibits a decreasing tendency until almost $R_1=0.4a_0$ value for $T_s=0.2a_0$. This decreasing tendency proceeds until $R_1=0.9a_0$ for excited state. This treatments can be explained as follows: In smaller values of core radius, the electron density in the well region is larger than that in the core of the structure for both $Z=0$ and $Z=1$ cases. Hence, the attractive Coulomb potential of the impurity does not pull so much down the energy levels. The expanding core region, however, is to become much more effective on the decreasing of energy levels. After certain $R_1$ values, the binding energies start to increase rapidly until $R_1=0.9a_0$ for 1s state. In case of $1p$ state, this increase, ongoing until $R_1=1.2a_0$, is more smooth compared to that in ground state. This behavior indicates that the probability of electron tunneling to core region rises with increasing of the core radius. In this case, the impurity pulls the energy values more down, especially ground state energies as expected, and so, this circumstance causes an increase on the donor binding energy. With further increasing of the core radius, the binding energies decrease smoothly and go towards bulk values due to the vanishing of the confining effects on the electron. Although similar behaviors are also observed for $T_s=0.4a_0$ case, the binding energy is bigger and their increasing becomes too steep. As is well known, if the barrier thickness increases, the tunneling probability of an electron confined in the core region (the well region) to the well region (the core region) decreases. We can easily conclude that the shell thickness has an important effect on the binding energies.

\begin{figure}
\includegraphics[width=3.0in]{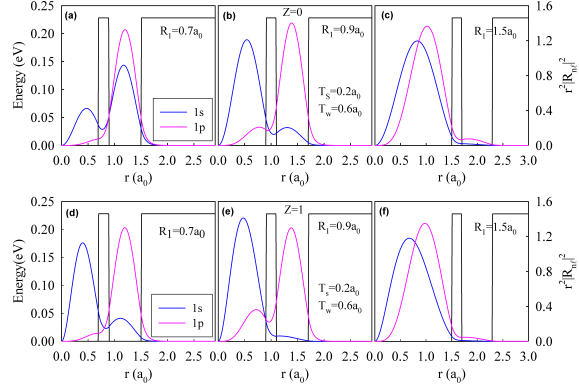}
\caption{\label{fig:9} Variation of the ground and excited states electron densities with r for constant shell thickness and well widths for $Z=0$ (top panel) and $Z=1$ (bottom panel) cases.}
\end{figure}

In order to make clear the impurity effects on the electron densities of ground and excited states, we plot Fig.\ref{fig:9} for different core radii and constant shell thickness and well widths. Figures \ref{fig:9} (a) and (d) show the electron densities in the same core radius for $Z=0$ and $Z=1$ cases, respectively. As seen from Fig.\ref{fig:9} (a), although the ground state density is substantially in the well region, the excited state density is almost completely in the well layer. When we compare Fig. \ref{fig:9} (d) with Fig.\ref{fig:9} (a), we can see that there is almost no effect of the impurity on the excited state density. However, the ground state density is seriously affected from the attractive coulomb potential of the impurity and as a result of this, the ground state density becomes localize substantially in the core region. We conclude that although the impurity has no serious effect on the excited state energies, it is much more effective on the ground states and it pulls the ground state energies more down. In Fig.\ref{fig:9} (b) and (e), it can be seen that the ground state densities become localize in the core region with increasing core radius for cases with and without the impurity. But, the strength of this localization is a bit more for $Z=1$ case, when it is compared to the $Z=0$ case. In addition, we can see that the impurity has somewhat effect on the excited states, even though it is lesser. In Fig.\ref{fig:9} (c) and (f), when the core radius $R_1=1.5a_0$, both ground and excited state densities are almost completely localize in the core region and the core/shell/well/shell QD structure exhibits a single core/shell QD properties with further increasing of the core radius.

\begin{figure}
\includegraphics[width=2.5in]{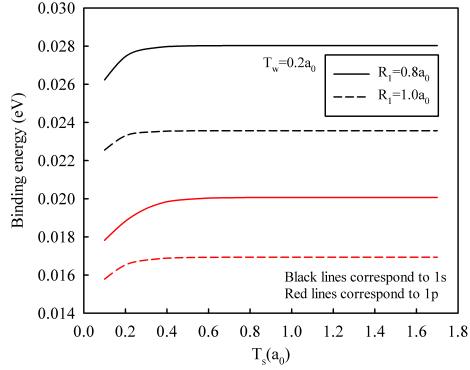}
\caption{\label{fig:10} Ground and first excited states impurity binding energy of a core/shell/well/shell QD as a function of the shell thickness ($T_s$) for $T_w=0.2 a_0$ and two different core radii, $R_1=0.8$ and $1.0 a_0$.}
\end{figure}

In multi-layered spherical quantum dot, the change in the barrier thickness is effective on the binding energy till a critical value of the thickness due to tunneling, which depends strongly on the barrier thickness, between the core and well region. The variation of the binding energy with the shell thickness is plotted in Fig.\ref{fig:10} for two different core radii ($R_1=0.8 $ and $ 1.0 a_0)$ and well width $T_w=0.2 a_0$. As seen from the figure, the binding energy of ground state increases until $T_s=0.3 a_0$ and it reaches up a certain value with further increasing of the $T_s$ for $R_1=0.8 a_0$. This increase is continuing up to $T_s=0.5 a_0$ for $R_1=1.0 a_0$. This is that because the core region becomes large, the electron feels lesser the effect of the well region. In excited states binding energies, these increases are observed until $T_s=0.2 a_0$ value for both $R_1=0.8$ and $1.0 a_0$. This behavior in binding energy can be interpreted as follows: In small shell thickness and when the well width is narrow, the electron has been confined into the core region. In addition to this, because the barrier thickness is thin, the effect of the well has been felt by electron. Therefore, as well as the impurity, the well has an influence on the energy levels and plays an important role in reducing the binding energy. As a result, in this shell thickness, the binding energy is small. If the barrier thickness exceeds a certain critical value, the electron in the core no longer feel the effect of well region and, therefore, the impurity binding energy remains constant with increasing shell thickness. Similar process is valid for excited state binding energies.

\begin{figure}
\includegraphics[width=2.5in]{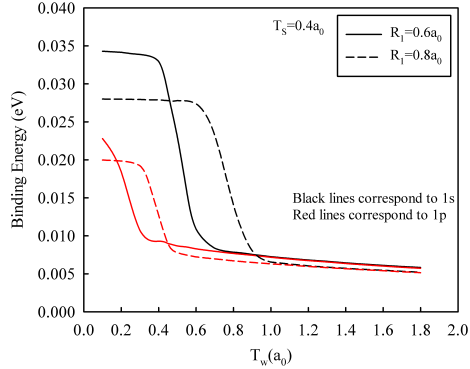}
\caption{\label{fig:11} Variation of the impurity binding energy of ground and first excited states with the well width ($T_w$) for $T_s=0.4 a_0$ and two different core radii, $R_1=0.6$ and $0.8 a_0$.}
\end{figure}

Figure \ref{fig:11} shows the variation of the donor binding energy with the well width for two different core radius ($R_1=0.6$ and $0.8 a_0$) and shell thickness $T_s=0.4 a_0$. If we look at the figure, the donor binding energies remain almost constant values for both ground and excited states (except for case with $R_1=0.6 a_0$) until certain well widths. After this certain values, these binding energies decrease sharply and approach to the same values. The physics of these behaviors can be explained as follows: When the well width is small, the electron becomes localize in the core region. Therefore, the impurity pulls more down the ground state energy and causes higher binding energies. With increasing well width, the electron in the core starts feeling the well region and while the probability distribution decreases in the core, that increases in the well. As a result of this, the binding energy decreases with increasing well width. If the width is increased further, the electron in the well does not feel the confinement effect of the walls and behaves like a free particle. Thus, both ground and excited states binding energies will become the free electron binding energies.

\section{Conclusion}
In this study, we have carried out the deeply investigation of the electronic properties and the donor binding energies of a core/shell/well/shell quantum dot structure. In order to determine the energy eigenvalues and corresponding wave functions, the Schr\"{o}dinger equation is solved numerically with the shooting method in the effective mass approximation for a finite confining potential. The effect of a hydrogenic donor impurity located at the center of the multi-layered spherical quantum dot is also considered . The results for cases with and without a hydrogenic impurity are comparatively presented as a function of the core radius and the layer thickness. We have concluded that the attractive Coulomb effect of the impurity results in a significant change of the electronic structure of the multi layered quantum dot. The size of the barrier, well and core layers are also effective to determine the electronic properties of the system. Trends of all results are in a good agreement with other theoretical studies. We hope that this study will be very useful and contribute to understand the electronic properties of a core/shell/well/shell spherical quantum dot with and without an impurity.

\section*{Acknowledgement}

This study is a part of M.Sc. Thesis prepared by H. Ta\c{s} at Physics Department of Selcuk University. One of the authors (M.S) thanks Selcuk University BAP office for their partial support.

\end{document}